\documentstyle[12pt]{article}
\def\Journal#1#2#3#4{{#1} {\bf #2}, #3 (#4)}

\def\NPB{{\em Nucl. Phys.} B}

\def\PLB{{\em Phys. Lett.} B}
\def\PRL{\em Phys. Rev. Lett.}
\def\PR{{\em Phys. Rev.}}
\def\PRA{{\em Phys. Rev.} A}
\def\PRB{{\em Phys. Rev.} B}
\def\PRD{{\em Phys. Rev.} D}

\def\RMP{{\em Rev. Mod. Phys.}}
\def\AP{{\em Ann. Phys.}}
\def\PREPC{{\em Phys. Rep.} C}

\def\AJP{{\em Am. J. Phys.}}
\def\JPA{{\em J. Phys.} A.}

\newcommand{\be}{\begin{equation}}
\newcommand{\ee}{\end{equation}}
\newcommand{\bea}{\begin{eqnarray}}
\newcommand{\eea}{\end{eqnarray}}

\newcommand{\hf} {{1\over2}}
\newcommand{\nonu}{\nonumber\\}
\def\ra{\rangle}
\def\la{\langle}
\def\s#1{{\bf#1}}
\def\tr{{\rm tr}}
\def\tell{\tilde\ell}

\begin{document}
\begin{center}
{\LARGE Renormalization Group Transformation for the Wave Function}\\
\vspace*{1cm}
Hanae El Hattab\footnote{elhattab@lpt1.u-strasbg.fr}$^a$,
Janos Polonyi\footnote{polonyi@fresnel.u-strasbg.fr}$^{ab}$\\
\vspace*{1cm}
$^a$Laboratory of Theoretical Physics, Louis Pasteur University\\
3 rue de l'Universit\'e 67087 Strasbourg, Cedex, France\\
\vspace*{.5cm}
$^b$Department of Atomic Physics, L. E\"otv\"os University\\
Puskin u. 5-7 1088 Budapest, Hungary\\
\vspace*{2cm}
{\LARGE Abstract}\\
\end{center}
The problem considered here is the determination of the hamiltonian of a first quantized
nonrelativistic particle by the help of some measurements of the location
with a finite resolution. The resulting hamiltonian
depends on the resolution of the measuring device. This dependence is
reproduced by the help of a blocking transformation on the
wave function. The systems with quadratic hamiltonian are studied in details.
The representation of the renormalization group in the space of observables
is identified.

\section{Introduction}
The renormalization group provides us a systematic method to study the 
dynamics of a subsystem which is embedded in a larger dynamical environment.
When the subsystem contains the degrees of freedom characterizing the physics
beyond the length scale of a certain ultraviolet 
cutoff, $\Lambda^{-1}$, then the $\Lambda$
dependence of the dynamics of the subsystem reveals  the 
dependence of the fundamental laws and the physical quantities on the scale of the 
observation, $\Lambda$. 

The cutoff can be implemented in a sharp or smooth
manner. There are furthermore two slightly different generic realizations of the sharp cutoff
depending on whether the degrees of freedoms are left intact by the
decomposition of the whole system into the subsystem and its environment.
What is usually employed for statistical or second quantized systems
\cite{kadanoff} \cite{wilsrg} is the factorization 
\be
{\cal H}={\cal H}_s\otimes{\cal H}_e\label{dprod}
\ee
of the Hilbert space of the whole system $\cal H$. The $\Lambda$ dependent
subspace ${\cal H}_s$ contains the states of the subsystem in question and the
states of the environment are in ${\cal H}_e$. With the proper choice of
${\cal H}_s$ one can ensure that the subsystem corresponds to usual 
canonical degrees of freedom. The typical implementation of this
factorization in Quantum Field Theory is when ${\cal H}_s$ is chosen to be
the space containing the multiparticle states where the energy or the
momentum of each particle is below $\Lambda$. 

Another, simpler method \cite{fesch} \cite{wegner} \cite{rau} better suited for
the first quantized systems, starts with the direct sum decomposition
\be
{\cal H}={\cal H}_s\oplus{\cal H}_e.\label{dsum}
\ee
There is now no general argument
to guarantee that the subspace ${\cal H}_s$ contains all state vectors
of some canonical degrees of freedom. In some important applications of this
procedure, such as the ground state and few low lying 
resonances in the study of the nuclear collisions \cite{fesch}, 
or the lowest Landau level in the case of the Quantum Hall Effect \cite{qhe}
${\cal H}_s$ consists of the states with (unperturbed) 
energy less than $\Lambda$.

The goal of the renormalization group is to identify the dynamics of the subsystem 
by means of the reduced time evolution operator
\be
U_s(t)={\rm Tr}_{{\cal H}_e}e^{-{i\over\hbar}tH},\label{efpr}
\ee
or
\be
U_s(t)={\cal P}_{{\cal H}_s}e^{-{i\over\hbar}tH}{\cal P}_{{\cal H}_s},\label{efsu}
\ee
in the case of the direct product or direct sum decomposition, respectively.
Here ${\rm Tr}_{{\cal H}_e}$ stands for the trace in the space ${\cal H}_e$ 
and ${\cal P}_{{\cal H}_s}$ denotes the projection operator into the subspace
${\cal H}_s$. The characteristic feature of such a projected dynamics is that the 
time evolution is not unitary because the state vectors
in ${\cal H}_s$ leak into ${\cal H}_e$. This nonunitarity has been confirmed
in nuclear physics where the name "optical potential" was coined to describe
its effects. 

The physical difference between (\ref{dprod}) and (\ref{dsum}) is that each
state in ${\cal H}_s$ has a "contamination" from ${\cal H}_e$ in the case of
the direct product. In fact, the low energy particle state in ${\cal H}_s$ has 
the cloud of the
virtual high energy particles from ${\cal H}_e$ in any interactive Quantum 
Field Theory. The mathematical origin of this problem is that the 
elimination of a degree of freedom 
converts the pure states into mixed ones. This issue can be better
understood in the renormalization of Quantum Field Theory  where the 
effective theory in ${\cal H}_s$ contains nonlocal vertices up to the distance
scale $O(\Lambda^{-1})$ which are described by the higher derivative terms
in the effective action. These terms prevent the construction
of the transfer matrix and the effective hamiltonian in ${\cal H}_s$.
As a result the specification of the initial and
final pure states of the particles in ${\cal H}_s$ is not sufficient to obtain
the transition amplitude. So long as the higher order derivative terms
are irrelevant \cite{jochen} \cite{herve}, this problem is naturally negligible. 

The sharp cutoff procedure is clearly an idealization, the actual separation
is defined by the interactions between the subsystem and its environment.
In fact, the higher energy states decouple from the subsystem in a continuous
rather than a sharp manner. The smooth cutoff procedure attempts to
introduce such a more natural decomposition. This is usually achieved
by means of a smearing function which is to make up the assumed gradual
decoupling of the short distance modes from the long distance ones.
In Quantum Field Theory this happens by the introduction of the
blocked field variable
\be
\phi(x)\longrightarrow\phi'(x)=\int dy\chi(x-y)\phi(y).\label{blcsp}
\ee
The effective action, $S'[\phi']$, is defined by the path integral
\be
e^{{i\over\hbar}S'[\phi']}=\int D[\phi(x)]\prod_{x}
\delta(\phi'(x)-\int dy\chi(x-y)\phi(y))e^{{i\over\hbar}S[\phi]}.
\ee
The periodic boundary condition for the field variable in its time argument
implements the trace in (\ref{efpr}). 

A natural choice for the smearing function,
\be
\chi_a(x)=\Omega^{-1}_d(a)\Theta(a-|x|),
\ee
where $\Omega^{-1}_d(a)$ is the volume of the sphere of radius $a$ 
in d-dimensions, corresponds to a sharp cutoff, $a$, in real-space. Another simple 
choice is
\be
\chi_\Lambda(x)=\int_{|q|\le\Lambda}{d^dq\over(2\pi)^d}e^{iq\cdot x},
\ee
which introduces a sharp cutoff in the momentum space.
In fact, (\ref{blcsp}) yields the blocking transformation
\be
\tilde\phi(q)\longrightarrow\tilde\phi'(q)=\tilde\chi(q)\tilde\phi(q)\label{blmp}
\ee
for the Fourier transform 
\be
\tilde\phi(q)=\int{d^dq\over(2\pi)^d}e^{iq\cdot x}\phi(x),
\ee
where
\be
\tilde\chi_\Lambda(q)=\Theta(\Lambda-|q|).
\ee
A cutoff which is smooth both in real and momentum space is given by 
\be
\tilde\chi_\Lambda(q)=e^{-{q^2\over\Lambda^2}}.
\ee

It is worthwhile noting that the 
procedures are similar in Minkowsky and Euclidean space-time except 
that the regulator is usually not Lorentz invariant
\footnote{The non-compactness of the Lorentz
group makes the cutoff nonrelativistic in a positive definite Hilbert space.
The negative norm states allows Lorentz invariant regularization but there is no
path integral representation for the transition amplitudes.}.
Another characteristic feature of the real-time blocking is that the complex phase factor lends
a singular dependence to the running coupling constants on the cutoff \cite{polo}.

The smooth cutoff can be implemented in (\ref{efsu}) by the replacement of
the projection operator ${\cal P}_{{\cal H}_e}$ by another operator whose 
eigenvalue is a smooth function of the momentum or the energy.
The price is the loss of the property ${\cal P}^2_{{\cal H}_e}={\cal P}_{{\cal H}_e}$.

The goal of this paper is to provide a "softening" of the
sharp cutoff (\ref{dsum}). This allows one to follow the resolution dependence of the
dynamics for the first quantized systems in a more realistic and systematic manner.
With a slight generalization of (\ref{efsu}) we introduce
\be
U_{s,s'}(t)=e^{-{i\over\hbar}tH_{s,s'}(t)}=
{\cal P}_{{\cal H}_s}e^{-{i\over\hbar}tH}{\cal P}_{{\cal H}_{s'}}\label{gefsu}
\ee
which describes the evolution of the system {\em from} the subspace ${\cal H}_{s'}$
{\em to} ${\cal H}_s$ and change the sharp cutoff of (\ref{dsum}) to a smooth
one in a manner reminiscent of (\ref{blcsp}). In realistic situations, the smearing
(\ref{blcsp}) is supposed to account for the effects of the short 
distance processes such as virtual particle emissions and absorptions. These processes
"dress up" the particles and modify their interactions. This appears as the evolution
of the coupling constants as the functions of the cutoff. We shall emulate
this regrouping of the dynamics from the degrees of freedom into the choice of the
coupling constants in first quantized Quantum Mechanics. 

Suppose that we measure (up to a global phase) the wave function of a particle that 
propagates in an unknown potential by an experimental device with space resolution 
$\Delta x=\sigma$. How does the potential, $V_\sigma(x)$, reconstructed by the help 
of these measurements depend on the space resolution? There is certainly
such a dependence
because the potential is smeared out and is slowly varying within the distance $\sigma$. 
We shall try to take the 
deformations of the propagation caused by the measuring device, such as the 
interference or decoherence, into account by a smearing of the wave 
function rather than by going into the second quantized description. One hopes
that if the smearing of the wave function was chosen appropriately then
such a simplification still contains the essential elements of the physics.

Our procedure can be summarized in a formal manner as follows.
In the Schr\"odinger functional formalism the Quantum Field Theory for the scalar field
$\phi(\s x,t)$ is considered as first quantized quantum mechanical system by converting 
$\phi(\s x,t)$ into coordinate, i.e. changing 
the internal space of the field variable into the external space of the 
quantum system. The
dimension of the latter is the number of degrees of freedom of the field theory
model, the number of points in the coordinate or the momentum space. 
The blocking (\ref{blmp}) can be viewed as a rescaling, or a suppression
of the coordinates $\phi(\s p,\omega)$ which depends on the choice of the
type of the coordinate, $(\s p,\omega)$. It drives the system into a lower dimensional
subspace. Our blocking in the first quantized formalism will be realized by
a smearing function $\chi(\s x-\s y)$ for the wave function which depends on the 
location in space. This would have been a $\phi(\s x,t)$-dependent rescaling
in the case of the field theory.

In Section 2. of the paper the blocking,
in particular the decimation in time, is reviewed for Quantum Mechanics. A simple
motivation of our blocking prescription in space is presented in Section 3. This
procedure is applied to different systems with quadratic hamiltonians in Section 4.
Section 5 is devoted to the construction of the representation of the renormalization
group in the space of observables. Finally, a brief summary of our results
is presented in Section 6.

\section{Renormalization Group in Quantum Mechanics}
We start by reviewing the motivation of using the renormalization group in
Quantum Mechanics. As in Quantum Field Theory, the idea of the renormalization
group appeared first for systems with ultraviolet divergences. For hamiltonians 
with a power-like or Dirac-delta type singular potential, the regularization and 
the subsequent renormalization was applied in \cite{sing}.
A partial resummation of the perturbation expansion by the introduction of the running
coupling constant was performed in \cite{run}. The singular 
quantum propagation inspired the implementation of the renormalization
group as a device to trace down the dependence on the time scale of the observations
in \cite{polo}. To understand this latter issue better, consider the wave function 
\be
\psi(x,t;y)=\la x|e^{-{i\over\hbar}tH}| y\ra=
\biggl({m\over2\pi i\hbar t}\biggr)^{1/2}e^{{im\over2\hbar t}(x-y)^2},
\ee
for a free particle in one dimension which satisfies the initial condition
\be
\psi(x,0;y)=\delta(x-y).\label{inic}
\ee
It keeps the relative phase difference between the point $y$ and the sphere with radius 
\be
\ell(t)=\sqrt{2t\hbar\phi\over m}
\ee
around $y$. The phase velocity,
\be
v_\phi=\sqrt{\hbar\phi\over2mt},\label{divvel}
\ee 
associated to this propagation has an ultraviolet, i.e. $t\to0$, divergence. 
This is in agreement
with the observation that the Wiener measure is concentrated on nowhere differentiable
paths and the velocity obtained form the trajectories of the path integral in
Quantum Mechanics diverges as (\ref{divvel}) \cite{schulm}. Such a surprising deviation 
from the analiticity of the classical trajectories suggests that the dependence of the result
of a measurement on the time of the observation may be unusual in Quantum Mechanics.

Based on a formal similarity between path integrals for a 0+1 dimensional Quantum Field Theory
and Quantum Mechanics, certain elements of the renormalization group method
can be replanted from Quantum Field Theory to Quantum Mechanics. The evolution
of the observables during the increase of the time scale of the measurement
was studied in Ref. \cite{polo} by means of the decimation of the 
integral variables of the path integral formalism. The concept of the running 
coupling constant was introduced by parametrizing the logarithm of the propagator 
\be
\la x|e^{-{i\over\hbar}\Delta tH}| y\ra=e^{{i\over\hbar}S(x,y;\Delta t)},
\label{action}
\ee
by the quantum action,
\be
S(x,y;t)=\sum_{n,m}g_{n,m}(t)(x-y)^n\biggl({x+y\over2}\biggr)^m,
\ee
in formal analogy with Quantum Field Theory. Here $g_{n,m}(t)$ is the coupling constant
which is a
periodic function of time for a quadratic, hermitean hamiltonian and the decimation in
real time creates singularities and zeros as a function of $t$. 
At the tree level one finds that the coupling constant 
$g_{n,m}(t)$ is relevant, i.e. is an increasing function of $t$ for 
\be
n\le2.\label{relt}
\ee 

It is important to keep in mind that the path integral may have 
ultraviolet divergences in Quantum Mechanics. 
We usually call a coupling constant renormalizable if the insertion of
the corresponding vertex into a graph reduces the degree of the overall divergence.
According to the power counting argument, $g_{n,m}(t)$ is renormalizable for
\be
\omega_{n,m}\equiv{m-n\over2}+1\ge0.\label{renormt}
\ee 
This comes from the expression
\be
\omega(G)=D+{2-D\over2}E-\sum_v\omega_v\label{prdiv}
\ee
for the overall divergence of a graph in $D$-dimensional Quantum Field Theory
with $E$ external legs. The summation is over the vertices $v$ and $\omega_v$
denotes the energy dimension of the vertex
\be
\omega_v\equiv\omega_{n,m}=D+m-{D\over2}(n+m).
\ee
Eq. (\ref{prdiv}) shows that the degree of divergence of the graph 
is not increased by the insertion of a renormalizable vertex.

Note that the discrepancy between (\ref{relt}) and (\ref{renormt}) indicates
that the class of relevant and marginal operators does not agree with the
class of the perturbatively renormalizable operators. 
It is not so surprising that the operators
with $n>2$ and $m>n+2$ are irrelevant and renormalizable in the same time
because the ultraviolet divergences are weaker in low dimensions. The other
type of discrepancy, which is in particular represented by the coupling constant 
$g_{n,m}$ with $(n,m)=(2,0)$, the kinetic energy, is more worrisome. But we believe that
the solution of this puzzle will come from a better understanding of the
renormalizability below the lower critical dimension, $D<2$. Observe
that in such dimensions the overall degree of divergence of a graph is
increased by adding external legs.
Thus there are infinitely many divergent Green functions and the
usual BPHZ procedure is not sufficient to remove the cutoff and to 
renormalize the theory. The more general nonperturbative arguments,
which assert the equivalence of the relevant and marginal operators 
with the renormalizable ones, suggest that similar result will only 
be established after a more carefully repetition of 
the renormalization program for $D<2$.

Furthermore, the irrelevant coupling constants $g_{n,m}$ with $n>2$ 
allow the construction of the 
"effective theories" which are non-renormalizable quantum systems with path integral 
description but without hamiltonian. This may happen because on the one hand, the
path integral is well defined for $\Delta t\not=0$ after the introduction of
the counterterms to remove the divergences, and on the other hand,
the Schr\"odinger equation is obtained in the limit $\Delta t\to0$ which does not 
exist for a non-renormalizable system.

We present in this paper a renormalization group transformation in the
coordinate space. The corresponding renormalized trajectory maps out 
the dependence of the measurements on the resolution of the measuring devices 
in the coordinate space. The renormalized trajectories for blocking
in space or time are actually unrelated for nonrelativistic systems.

\section{Measurements and Blocking}
The implementation of the smearing of the wave function as a method to
separate the subsystem from its environment can be motivated by inspecting
a typical measurement process.

{\it Measurements with finite resolution:}
Suppose that we want to determine the location of a particle. The devices,
such as the photosensitive film, the steamer chamber or the solid 
state detectors are
based on certain interactions between the particle in question and the material of the
detecting device. As far as the latter is concerned, it is useful to distinguish
two length scales. One, the total length, $\ell_t$, is the size of the volume element
where the detector is sensitive. Another one, $\ell_c<<\ell_t$, 
is the coherence length, the size of the microscopic structure used in the detection
process. This is the size of the atom or the cluster of atoms which enters into 
interaction with our particle. One has to average the
quantum amplitude on the scale $x<\ell_c$ and the probability for $\ell_c<x<\ell_t$
\footnote{One encounters similar treatment in the description of the scattering processes
where one sums the amplitude for the initial states and the probability for the final states.}.
It is assumed here that the motion of the particle is coherent within the whole detector,
i.e. the soft particle emission and other sources of decoherence are neglected and the
coherence of the measurement is lost for $x>\ell_c$ due to the components of the detector.
We introduce the projection operator,
\be
P(x)=|\chi_x\ra\otimes\la\chi_x|,
\ee
$P^2(x)=P(x)$, where $|\chi_x\ra$ denotes the state of the microscopic
part of the measuring device located at $x$. 
$P(x)$ corresponds to the measurement process performed by the 
microscopic structure located at the point $x$. 
The probability of finding the particle with the wave function $\psi(x)$ is approximated by
\be
{\cal P}_m=\int dx\rho(x)\tr P(x)|\psi\ra\otimes\la\psi|,
\label{measdm}
\ee
where $\rho(x)$ is to take into account the distribution of the microscopic
objects in the detector.

{\it Blocking:}
The formal relation with the Kadanoff-Wilson blocking is established
by writing (\ref{measdm}) as
\be
{\cal P}_m=\int dx\rho(x)|\psi_\chi(x)|^2,
\ee
where
\be
\psi_\chi(x)=B_\chi\psi(x)=\int dy\chi^*(x-y)\psi(y).\label{blwf}
\ee
The amplitude $\chi(x-y)$  defined as
\be
\chi(y)=\la y|\chi_0\ra
\ee
is the wave function of the microscopic part of the device
and is vanishing for $|y|\ge\ell_c$.
The linear operator $B_\chi$ introduced here performs the
smearing, i.e. the blocking of the wave function.
The sharp cutoff in one dimensional space is realized by the smearing function
\be
\chi_a(x)={1\over a}\Theta(a-|x|)\label{scco}
\ee
where $a\approx\ell_c$.

Our goal is the study of the dependence on the parameter $\ell_c$. This will be achieved
by finding the quantum mechanics of the particle, i.e. its hamiltonian $H_\chi$, 
which is reconstructed by the help of the measurements of the coordinates.
The dependence on the classical, incoherent part, $\rho(x)$ is trivial
from our point of view so we shall take $\rho(x)=\delta(x)$ in this work.

Note that the blocking operator, $B_\chi$, introduced in this way is not necessarily
unitary or even invertible. These features reflect our intention to loose
information during the blocking. In fact, the retaining of the infrared part of
a state decreases the norm of its wave function. The deficiency,
\be
1-\la\psi(t)|B_\chi^\dagger B_\chi|\psi(t)\ra,
\ee
will be a complicated, nonmonotonic function of time. It is useful
to introduce the properly normalized blocked wave function, 
\be
\psi_R(x,t)=\sqrt{Z(t)}\psi_\chi(x,t),
\ee
by the help of the wave function renormalization constant $Z$. 

{\it Gaussian smearing:}
Another blocking procedure is a Gaussian suppression in the momentum space,
\be
B(\sigma^2)=e^{-{\sigma^2_+\over2\hbar^2}p^2},\label{gausmi}
\ee
where $\sigma^2_\pm=\sigma^2\pm i\epsilon$ and $0<\epsilon\to0$. 
The normalized wave functions are generated by the nonlinear operator
\be
B_R(\sigma^2)=e^{-{\sigma^2_+\over2\hbar^2}p^2-V(\sigma^2)},
\ee
where $Z^{1/2}=e^{-V}$ satisfies the inequality
\be
\sigma^2{\partial\over\partial\sigma^2}V
=-\hf\sigma^2{\partial\over\partial\sigma^2}\log Z=
-{\la\psi_\sigma|{\sigma^2\over2\hbar^2}p^2|\psi_\sigma\ra
\over\la\psi_\sigma|\psi_\sigma\ra}\le0.\,
\ee
showing that the norm of the blocked wave function is decreased during 
blocking by the suppression of the higher momentum components of the states.
The transformation
\be
\tilde\psi_\sigma(p)=e^{-{\sigma^2_+\over2\hbar^2}p^2}\tilde\psi(p),
\ee
of the Fourier transform of the wave function $\tilde\psi(p)$ results in the blocking
\be
\psi_\sigma(x)={1\over\sqrt{2\pi\sigma^2}}\int dy
e^{-{1\over2\sigma^2_+}(x-y)^2}\psi(y)\label{blcspk}
\ee
for the wave function $\psi(x)$.

{\it Landau pole:} The renormalization in Quantum Field Theory 
consists of the successive elimination of degrees of freedom from
the theory. The effect of a mode under elimination is kept in the
system by modifying the dynamics, i.e. in the scale
dependence of the blocked hamiltonian for the retained modes.
We have a harmonic oscillator associated to each particle mode of the
theory. It may happen that the elimination of some modes makes
the quadratic part of the hamiltonian negative semidefinite. At this point,
at the Landau pole,
the perturbation expansion fails and the fluctuations become large.
This indicates that the new hamiltonian which includes the effects
of the eliminated modes, can not be written in the same functional
form as the original one, at least within the framework of 
perturbation expansion.
A Landau pole is called infrared or ultraviolet depending on the
way it is encountered during the renormalization procedure.

Quantum mechanical systems display singularities
in the function of the cutoff, as well, but they turn out to be the opposite 
of the Landau poles of the quantum field theory models in that the 
quantum fluctuations disappear there. This is because we make
the blocking in coordinate space in quantum mechanics which
corresponds to the internal space of the Quantum Field Theory
model and such a blocking modifies the noncommuting quadratic pieces
of the hamiltonian, kinetic energy and harmonic oscillator 
potential, differently. We find, in the case of the quadratic systems considered
here, that only the kinetic energy vanishes at the Landau pole leaving behind a 
harmonic oscillator potential.

The explicit blocking construction guarantees the existence of the blocked 
action so our Gaussian blocking form a semi-group,
\be
B(0)=1,~~~~~B(\sigma'^2)B(\sigma^2)=B(\sigma'^2+\sigma^2).\label{rgroup}
\ee
The ultraviolet Landau pole occurs at $\sigma_L$ if $B(\sigma^2_L)$ is not invertible. 
There is no infrared Landau pole in this formalism.

The inverse of the blocking transformation is
\be
B(\sigma^2)^{-1}=e^{{\sigma^2_-\over2\hbar^2}p^2}.\label{invbl}
\ee
In fact,
\bea
\la x|B(\sigma^2)|y\ra&=&\int{dp\over2\pi\hbar}
e^{{i\over\hbar}p(x-y)-{\sigma^2_+\over2\hbar^2}p^2}\nonu
&=&{1\over\sqrt{2\pi\sigma^2_+}}e^{-{1\over2\sigma^2_+}(x-y)^2},\nonu
\la x|B(\sigma^2)^{-1}|y\ra&=&\int{dp\over2\pi\hbar}
e^{{i\over\hbar}p(x-y)+{\sigma^2_-\over2\hbar^2}p^2}\nonu
&=&{1\over\sqrt{-2\pi\sigma^2_-}}e^{{1\over2\sigma^2_-}(x-y)^2},
\eea
which yields
\bea
\la x|B(\sigma^2)B(\sigma^2)^{-1}|y\ra
&=&\int dz\la x|B(\sigma^2)|z\ra\la z|B(\sigma^2)^{-1}|y\ra\nonu
&=&-{1\over2\pi\sigma^2}e^{{y^2\over\sigma_-^2}-{x^2\over\sigma_+^2}}\nonu
&&\times\int dze^{-{z^2\over2}({1\over\sigma_+^2}-{1\over\sigma_-^2})
+z({x\over\sigma_+^2}-{y\over\sigma_-^2})}\nonu
&=&{1\over\sqrt{4\pi i\epsilon}}
e^{-{1\over4i\epsilon}(x-y)^2}.\label{inverse}
\eea

$B(\sigma^2)^{-1}$ is well defined in the Hilbert space span by the wave 
functions
\be
\psi(x)=o\left(e^{-{1\over2\sigma^2}x^2}\right)\label{ascond}
\ee
for $x\to\infty$. If this asymptotic condition (\ref{ascond}) is not respected then the 
multiple integrals appearing in the matrix elements are not uniform convergent
indicating an ultraviolet Landau pole. This can clearly be seen in the case of 
the matrix elements of $B(\sigma^2)B^{-1}(\sigma^2)$ between the harmonic oscillator
eigenstates
\bea\label{inersme}
\la n|B(\sigma^2)B(\sigma^2)^{-1}| m\ra&=&{1\over\ell\sqrt{\pi}}
\int dxdye^{-{1\over2\ell^2}(x^2+y^2)}\\
&&\la x|B(\sigma^2)B(\sigma^2)^{-1}|y\ra
H_n\left({x\over\ell}\right)H_m\left({y\over\ell}\right),\nonumber
\eea
where $\ell^2=\hbar/m\omega$.
When the integration over $x$ and $z$ in second line of  (\ref{inverse})
is performed before the $y$ integration, then (\ref{inersme}) is always well
defined and finite. On the contrary, when the scalar product between $B(\sigma^2)^{-1}$
and the state $|m\ra$ is made for fixed values of $x$ and $z$,
then (\ref{inersme}) is divergent for $\sigma^2<\ell^2$. The blocking transformation
can not be inverted within the physical space of such a harmonic oscillator. 
We encounter an ultraviolet Landau pole because one can not 
create a state with the localization
$\Delta x<\ell$ in the Hilbert space span by the harmonic oscillator eigenstates.

The blocking is supposed to loose information and should be noninvertible
such as (\ref{scco}). Though the $i\epsilon$ modification of the Kadanoff-Wilson
blocking procedure guarantees the formal inverse in the complete
Hilbert space, it may happen that the blocking transformation
is not invertible in the subspace determined by the physical problem.

{\it Blocked dynamics:}
We shall introduce the running coupling constants in the hamiltonian and the lagrangian.
The time evolution of the original wave function
\be
i\hbar\partial_t\psi=H\psi,
\ee
induces 
\be
i\hbar\partial_t\psi_\chi=H_\chi\psi_\chi
\ee
as the equation of motion for the blocked wave function where
\be
H_\chi B_\chi=B_\chi H.\label{blham}
\ee
The hamiltonian that generates the time evolution for the blocked wave function
can be obtained directly,
\be
H_\sigma=B(\sigma^2)HB(\sigma^2)^{-1}=e^{-{\sigma^2_+\over2\hbar^2}p^2}
He^{{\sigma^2_-\over2\hbar^2}p^2},\label{hambl}
\ee
as long as the blocking is invertible. Notice that $H_\sigma$ is {\em not} the
hamiltonian sandwiched between wave packet states, it is defined instead as the generator
of the time evolution for the wave packets.
The way $H_\sigma$ is obtained corresponds to performing a diffusion process, using 
the {\em free} quantum propagation for
\be
\tau={\sigma^2 m\over\hbar}\label{imagt}
\ee
which is an imaginary time. It is interesting to notice that the choice
\be
B_\tau=e^{-{\tau\over\hbar}H}
\ee
correspond to a special blocking which generates the time evolution
\be
\psi_\chi(x,t)=e^{-{i\over\hbar}(t-i\tau)H}\psi(x,0).
\ee

Starting with the form $H={p^2\over2m}+V(x,p)$ the blocking transformation
can be written as
\bea
H_\sigma&=&e^{-{\sigma^2_+\over2\hbar^2}p^2}
\biggl({p^2\over2m}+V(x,p)\biggr)
e^{{\sigma^2_-\over2\hbar^2}p^2}\nonu
&=&{p^2\over2m}+V(x,p)+e^{-{\sigma^2_+\over2\hbar^2}p^2}
[V(x,p),e^{{\sigma^2_-\over2\hbar^2}p^2}].
\eea
The differential form of the renormalization group equation for the length scale 
independent operator $h=\sigma^{-2}H_\sigma$ is
\be
\sigma{\partial\over\partial\sigma}h=-2h+{\sigma^2\over\hbar^2}[h,p^2].
\ee
The term $H_{n,m}(x,p)$ which is an $n$-th and $m$-th order homogeneous
function in $p$ and $x$, respectively, is represented by the length scale independent form,
\be
h_{n,m}(x,p)=\sigma^{2-n-m}H_{n,m}(x,p)
\ee
in the hamiltonian. The action of an infinitesimal renormalization 
group transformation is
\be
{\cal L}h_{n,m}=\sigma{\partial\over\partial\sigma}h_{n,m}=(2-n-m)h_{n,m}
+{\sigma^2\over\hbar^2}[h_{n,m},p^2].\label{rgecf}
\ee
The scaling operators satisfy the equation
\be
{\cal L}h_{n,m}=\nu_hh_{n,m}.
\ee
A simple class of the scaling operators is made up by
the powers $p^n$, with critical exponent $\nu=2-n$. The only
relevant operators of this class come from a homogeneous vector 
potential and the kinetic energy. Higher powers of the momentum, 
like the relativistic corrections are irrelevant. 

The quantum action, $S(x,y;t)$, introduced in (\ref{action}) is the
starting point for the space and time dependent running coupling constants of the lagrangian. 
The blocking transformation for $S(x,y;t)$ is given by
\bea
e^{{i\over\hbar}S(x,y;t)}&=&\la x|e^{-{i\over\hbar}tH}|y\ra\nonu
&\longrightarrow&\la x|B(\sigma^2)e^{-{i\over\hbar}tH}B(\sigma^2)^{-1}|y\ra\nonu
&=&\la x|e^{-{i\over\hbar}tH_\sigma}|y\ra\nonu
&=&e^{{i\over\hbar}S_\sigma(x,y;t)}.
\eea
For the generalized problem in (\ref{gefsu}) we introduce
\be
e^{{i\over\hbar}S_{\sigma,\sigma'}(x,y;t)}
=\la x|B(\sigma^2)e^{-{i\over\hbar}tH}B(\sigma'^2)^{-1}|y\ra.\label{genact}
\ee
The quantum action $S_{\sigma,\sigma'}(x,y;t)$ describes the propagation of a state
with localization $\sigma'$ at $y$ into another one with localization $\sigma$ 
at $x$. We shall consider only the case $\sigma'=0$ for simplicity. 

{\it Semiclassical limit:}
The commutator in the right hand side of (\ref{rgecf}) is $O(\hbar)$
and the nonclassical, second term is thus $O(\hbar^{-1})$ indicating
that the anomalous dimensions diverge in the semiclassical limit.
This can be traced back to the factor $\hbar^{-2}$ in the exponent
of (\ref{gausmi}). 

The dimensional parameter of Quantum Mechanics has two different roles 
in this formalism. One is related to 
its usual apparence in the time evolution operator, the other is through its
presence in the blocking, (\ref{gausmi}). The divergence of the anomalous
dimension (\ref{rgecf}) originates from the second role, the divergence
of the diffusion time (\ref{imagt}) what realizes the blocking. In other
words, any finite length scale $\sigma^2$ becomes large compared to $\hbar$
when the latter is assumed to approach zero. In order to decouple the two
different roles we can modify (\ref{gausmi}) as
\be
B(\sigma^2)=e^{-{\sigma^2_+\over2\hbar^2_0}p^2},\label{gausmim}
\ee
by introducing a new constant, $\hbar_0$, with the same dimension as $\hbar$ which
remains unchanged in the semiclassical limit. This amounts to the replacement
\be
\sigma^2\longrightarrow{\hbar^2\over\hbar^2_0}\sigma^2
\ee
in the expressions obtained so far and makes the anomalous dimension 
equivalent with the classical dimension in the semiclassical limit,
as expected.
We shall return to the contribution of the last term of (\ref{rgecf}) in 
Section 5.

\section{Quadratic Hamiltonians}
We now apply the blocking procedure introduced above for quadratic systems and obtain
their renormalized trajectory.

\subsection{Free Particle}
The blocked wave function is 
\bea
\psi_\sigma(x,t;y)&=&e^{{i\over\hbar}S_{\sigma,0}(x,y;t)}\nonu
&=&\sqrt{m\over2\pi i\hbar t}\int {dz\over\sqrt{2\pi\sigma^2}}
e^{-{1\over2\sigma^2}(x-z)^2+{im\over2\hbar t}(z-y)^2}\nonu
&=&\sqrt{m(\sigma,0)\over2\pi i\hbar t}
e^{{im(\sigma,0)\over2\hbar t}(x-y)^2},\label{freebwf}
\eea
where
\be
\ell^2={t\hbar\over m}
\ee
and the mass parameter is defined by
\be
m(\sigma,0)=m{1\over1-i{\sigma^2\over\ell^2}}.
\ee
The probability density of the blocked wave function is
\be
|\psi_\sigma(x,t;y)|^2={\cal N}^{-1}e^{-{1\over\Delta x^2}(x-y)^2},
\ee
with
\bea
\Delta x^2&=&\ell^2\biggl({\ell^2\over\sigma^2}+{\sigma^2\over\ell^2}\biggr),\nonu
{\cal N}&=&2\pi\ell^2\sqrt{1+{\sigma^4\over\ell^4}}.
\eea
The norm of the blocked wave function gives
\be
Z=2\sqrt{\pi}\sigma.
\ee

It is easy to understand that $\Delta x^2$ reaches its minimum at $\sigma=\sigma_{cr}=\ell$. 
The original wave function has the period length $2\pi\ell^2$ in function
of the length square of the propagation. For the under-smeared case with
$\sigma<<\sigma_{cr}$ the destructive interference is building up 
in  (\ref{freebwf}) as we increase $x-y$ when the phase of the original
wave function changes by $\pi$ within the interval $[x+\sigma,x]$,
\be
{(x+\sigma-y)^2-(x-y)^2\over\ell^2}\approx\pi,
\ee
which yields 
\be
\Delta x=x-y\approx{\pi\over2}{\ell^2\over\sigma}-{\sigma\over2}\approx
{\ell^2\over\sigma}.\label{uvsing}
\ee
For $\sigma>>\sigma_{cr}$ we have over-smearing because the interference 
is now as destructive as it can be even within the longest period length 
at $x-y\approx0$. But since the integration is done in the distance while
the periodicity is in the distance square the cancellation within a period
length is more complete for short period length, i.e. $|x-y|>>\ell$. The
dominant contribution coming from $x-y\approx0$ is thus suppressed
without any further interference simply by the smearing,
\be
\Delta x^2\approx\sigma^2.
\ee
The two opposite effects are balanced at the crossover $\sigma=\sigma_{cr}$.

The effective mass read off from the lagrangian or hamiltonian is
renormalization group invariant since $[H,B(\sigma^2)]=0$. The translation
invariant propagation between states with the same smearing,
$S_{\sigma,\sigma}(x,y;t)$, is scale invariant, $m(\sigma,\sigma)=m$.

\subsection{Harmonic Oscillator}
{\it Wave function:} The quantum action,
\bea
{i\over\hbar}S(x,y;t)=\ln\la x|e^{-{i\over\hbar}tH}|y\ra
&=&{i\over2}\cot t\omega{x^2+y^2\over\ell^2}
-{i\over\sin\omega t}{xy\over\ell^2}\nonu
&&-\hf\ln2\pi i\ell^2\sin\omega t,\label{holag}
\eea
where
\be
\ell^2={\hbar\over m\omega},
\ee
corresponds to the hamiltonian
\be
H={p^2\over2m}+{m\omega^2\over2}x^2
={p^2\over2m}+{\hbar\omega\over2}{x^2\over\ell^2}.
\ee
The blocked wave function is then given by
\bea
\psi_\sigma(x,t;y)&=&e^{{i\over\hbar}S_{\sigma,0}(x,y;t)}\nonu
&=&{1\over\sqrt{4\pi^2i\ell^2\sigma^2\sin\omega t}}
\int dze^{-{1\over2\sigma^2}(x-z)^2-{a\over2}(z^2+y^2)-bzy}\nonu
&=&{1\over\sqrt{2\pi i\rho^{-1}\ell^2\sin\omega t}}
e^{-\rho[{a\over2}(x^2+y^2)+bxy+\hf\sigma^2(a^2-b^2)y^2]},
\eea
where
\be
a=-{i\over\ell^2}\cot\omega t,~~~b={i\over\ell^2\sin\omega t},~~~
\rho={1+i{\sigma^2\over\ell^2}\cot\omega t
\over1+{\sigma^4\over\ell^4}\cot^2\omega t}.
\ee
The probability distribution is
\be
|\psi_\sigma(x,t;y)|^2={\cal N}^{-1}
e^{-{1\over\Delta^2x}(x-y/\cos\omega t)^2},
\ee
with
\bea
\Delta^2x&=&\ell^2\biggl({\ell^2\over\sigma^2}\tan^2\omega t+{\sigma^2\over\ell^2}\biggr),\nonu
{\cal N}&=&2\pi\ell^2\sin\omega t\sqrt{1+{\sigma^4\over\ell^4}\cot^2\omega t}.
\label{hohfp}
\eea
Note that the time reversed process starts with a wave packet and 
the $y$ dependence describes the oscillation of its center. 
The trigonometric functions in our expressions are due to the multiple
reflection from the potential. 

The width of the distribution in $x$ is minimal, $\Delta x=\sigma$, for focusing
or anti-focusing, 
\be
t\approx\cases{nT&focusing,\cr\bigl(n+\hf\bigr)T&anti-focusing,\cr} 
\ee
where $T=2\pi/\omega$. The localization is very strong after blocking 
for small $\sigma$ at focusing or anti-focusing because the original wave function
is perfectly localized. The crossover at
\be
\sigma_{cr}=\ell\sqrt{|\tan\omega t|}
\ee
indicates that the system is over-smeared close to focusing or anti-focusing 
and under-smeared otherwise. We found
\be
Z=2\sqrt{\pi}\sigma|\cos\omega t|
\ee
for the wave function renormalization constant. The norm of the blocked wave
function, $Z^{-1/2}$, shows minima for focusing and anti-focusing when the
wave function is well localized, i.e. is concentrated in the ultraviolet regime.

{\it Hamiltonian:}
One can easily find the blocked version of the hamiltonian, satisfying the relation
\be
H_\sigma e^{-{\sigma^2\over2\hbar^2}p^2}=e^{-{\sigma^2\over2\hbar^2}p^2}
\biggl({p^2\over2m}+{m\omega^2\over2}x^2\biggr).
\ee
The direct computation gives
\be
H_\sigma={p^2\over2m(\sigma)}+{m(\sigma)\omega^2\over2}x^2
-{m(\sigma)\omega^2\sigma^4\over2\ell^4}x^2
+i{\omega\sigma^2\over2\ell^2}(xp+px),\label{blhho}
\ee
with
\be
m(\sigma)=m{1\over1-{\sigma^4\over\ell^4}}.\label{hormass}
\ee
The coefficient of $x^2$ turns out to be renormalization group invariant. 
Nevertheless it is reasonable to keep the frequency independent of $\sigma$
because the blocking in space can not influence the period length in time, the quantities
without length dimension, in general. The effective mass has a singularity at  $\sigma=\ell$
and is negative for $\sigma>\ell$. The inverse of the naive blocking 
transformation (\ref{blcspk}) is well 
defined for $\sigma>\ell$ and the system has an ultraviolet Landau pole at
$\sigma=\ell$ where the inverse of the coefficient of the kinetic energy, 
$m(\sigma)$, diverges. The movement of the particle is more and  more difficult to
observe as space resolution, $\sigma$, approaches the localization length, 
$\ell$, from below. The motion disappears completely at $\sigma=\ell$ 
because the localisation of the particle happens to be compensated for
exactly by the smearing. 

The result of the direct computation of the blocked hamiltonian 
\be
H_\sigma={p^2\over2m}+V(x)+e^{-{\sigma^2_+\over2\hbar^2}p^2}
[V(x),e^{{\sigma^2_-\over2\hbar^2}p^2}],
\ee
performed at the infrared side of the Landau pole
reproduces (\ref{blhho}). This demonstrates that the $i\epsilon$ prescription
makes the blocking transformation (\ref{invbl}) invertible and the Landau pole
of the naive blocking procedure disappears.

{\it Action:} The transformation of the quantum action under blocking is
\bea
e^{-{a\over2}(z^2+y^2)-bzy}&\longrightarrow&
{1\over\sqrt{-4\pi^2\sigma^4}}\int dz_1dz_2\\
&&\times e^{-{1\over2\sigma^2_+}(x-z_1)^2-{a\over2}(z^2_1+z^2_2)
-bz_1z_2+{1\over2\sigma^2_-}(z_2-y)^2}.\nonumber
\eea
The gaussian integral results
\bea
{i\over\hbar}S_\sigma(x,y;t)&=&
-{1\over1-{\sigma^4\over\ell^4}}[{a\over2}(x^2+y^2)+bxy
-{\sigma^2\over2\ell^4}(x^2-y^2)]\nonu
&&-\hf\ln{2\pi i\ell^2(1-{\sigma^4\over\ell^4})\sin\omega t}
\eea
which is the logarithm of the transition amplitude for the wave packets. 
The unitary part of the propagation can be obtained from (\ref{holag})
by replacing the mass by its running value (\ref{hormass}).
The probability density,
\be
\left|e^{{i\over\hbar}S_\sigma(x,y;t)}\right|^2
={\cal N}^{-1}e^{-{1\over\Delta x^2}(x^2-y^2)},
\ee
with
\bea
\Delta x^2&=&\ell^2\biggl({\sigma^2\over\ell^2}-{\ell^2\over\sigma^2}\biggr),\nonu
{\cal N}&=&\left|2\pi\ell^2\left(1-{\sigma^4\over\ell^4}\right)\sin\omega t\right|,
\eea
shows the presence of the Landau pole at $\sigma=\ell$ where the sign of the
kinetic term changes. The time dependence drops from the blocking transformation
because both the initial and the final states are smeared in the same manner. 

The blocked action yields normalizable wave functions only for the over-smeared case
where the localization of the blocked states is weaker then 
the localization supported by the dynamics. In the operator formalism
the usual positive-mass hamiltonian was obtained on the ultraviolet
side of the Landau pole because in that case the energy, rather than
the states with a certain localization, was followed.

Note that the piece $1-\sigma^4/\ell^4$ in the quantum action can be interpreted as
a multiplicative factor to $\hbar$. As we approach the Landau pole the 
path integral becomes dominated by the vicinity of the classical
trajectory and we recover the semiclassical limit, the center of the 
blocked wave packets follow the classical equation of motion. The absence
of the quantum fluctuations at the Landau pole can be seen in the operator 
formalism, too. There the vanishing of the kinetic energy reduces the
quantum problem into a classical one at the Landau pole.

The lesson of this simple system is that the system possesses a
Landau pole where the localization length agrees with the block size.
The effective, running parameters of the hamiltonian (\ref{blhho}) change
slowly and stay reasonable so long as $\sigma<<\sigma_L=\ell$. The fundamental change 
of the dynamics around the Landau pole, $\sigma\approx\ell$, is due to 
the fact that the particle 
is better localized than the space resolution of the observation for $\sigma>>\sigma_L$. 
One looses sight of the motion and the particle propagation appears anomalous
at the infrared side of this crossover. Thus the 
sudden changes or singularities in the mass
indicates that the localization length is passed by the observational scale.

\subsection{Electron in Homogeneous Magnetic Field}
The free particle gave renormalization group invariant dynamics and the
harmonic oscillator yielded a nontrivial renormalized trajectory with Landau
pole. These different behaviors suggest the investigation of the propagation
of an electron in homogeneous magnetic field.
We start with the two dimensional hamiltonian in the presence of the 
background field $(A_x,A_y)=(0,Bx)$,
\be
H={1\over2m}p_x^2+{1\over2m}\left(p_y-{eB\over c}x\right)^2
\ee
whose eigenfunctions are
\be
\psi_{p,n}(x,y)={1\over\sqrt{\pi^{1/2}\ell2^nn!}}e^{{i\over\hbar}p_yy}
H_n\left({x-x_p\over\ell}\right)e^{-{(x-x_p)^2\over2\ell^2}},
\ee
with the eigenvalues
\be
E=\hbar\omega_L\left(n+\hf\right),
\ee
where
\be
x_p={cp_y\over eB},~~~\omega_L={|e|B\over mc},~~~\ell^2={\hbar\over m\omega_L}.
\ee
The corresponding quantum action can be written as
\bea
{1\over\hbar}S(\s u,\s v;t)&=&{1\over2\tell^2}(\s u-\s v)^2
+{1\over2\ell^2}(u_x+v_x)(u_y-v_y)\nonu
&&+i\log4\pi i\ell^2\left|\sin{\omega_Lt\over2}\right|,
\eea
by means of the notation
\be
\tell^2=2\ell^2\tan{\omega_Lt\over2}.
\ee

{\it Wave function:}
A possible gauge covariant generalization of the blocked wave function is
\be
\psi_\chi(\s u)=B_\chi[\s A]\psi(\s u)=\int d\s v\chi^*(\s v)
e^{{ie\over\hbar c}\int_{\s u+\s v}^{\s u}d\s w\s A(\s w)}\psi(\s u+\s v),\label{gcblwf}
\ee
where the integration in the exponent is performed along the straight
line connecting $\s u+\s v$ and $\s u$. This definition yields
\be
\psi_\sigma(\s u)=\int{d\s v\over2\pi\sigma^2} 
e^{-{1\over2\sigma^2}(\s u-\s v)^2+{ie\over\hbar c}\int_{\s v}^{\s u}d\s w
\s A(\s w)}\psi(\s v)\label{gism}
\ee
for the Gaussian blocking. Note that this blocking does not correspond any more
to a diffusion process, i.e. the propagation in imaginary time because the gauge phase 
factor is evaluated along a given path instead of summed over all trajectories.
The blocked action reads as
\bea
e^{{i\over\hbar}S_\sigma(\s u,\s v;t)}&=&-\int{d\s wd\s z\over(2\pi\sigma^2)^2}
\exp\biggl\{-{1\over2\sigma_+^2}(\s u-\s w)^2
+{ie\over\hbar c}\int_{\s w}^{\s u}d\s w'\s A(\s w')\nonu
&&+{i\over\hbar}S(\s w,\s z;t)+{1\over2\sigma_-^2}(\s z-\s v)^2
-{ie\over\hbar c}\int_{\s z}^{\s v}d\s z'\s A(\s z')\biggr\}.
\eea

We take the wave function
\be
\psi(\s u,t;\s v)=e^{{i\over\hbar}S(\s u,\s v;t)}
\ee
and perform the blocking,
\bea
\psi_\sigma(\s u,t;\s v)&=&e^{{i\over\hbar}S_{\sigma,0}(\s u,\s v;t)}\nonu
&=&\int{d\s w\over2\pi\sigma^2} 
e^{-{1\over2\sigma^2}(\s u-\s w)^2+{i\over2\ell^2}(u_x+w_x)(u_y-w_y)}
\psi(\s w,t;\s v)
\eea
with the result
\bea
\psi_\sigma(\s u,t;\s v)&=&\left(4\pi i\ell^2\left(1-i{\sigma^2\over\tell^2}\right)
\left|\sin{\omega_Lt\over2}\right|\right)^{-1}\nonu
&&\times\exp\left[{i\over2\tell^2}
{1+i{\sigma^2\tell^2\over4\ell^4}\over1-i{\sigma^2\over\tell^2}}(\s u-\s v)^2\right]\nonu
&&+{i\over2\ell^2}(u_x+v_x)(u_y-v_y).
\eea
The probability density is then 
\be
\left|\psi_\sigma(\s u,t,\s v)\right|^2
={\cal N}^{-1}e^{-{1\over\Delta x^2}(\s u-\s v)^2},
\ee
with
\bea
\Delta x^2&=&\ell^2\cos^2{\omega_Lt\over2}\left(4{\ell^2\over\sigma^2}\tan^2{\omega_Lt\over2}
+{\sigma^2\over\ell^2}\right),\nonu
{\cal N}&=&16\pi^2\ell^4\left(1+{\sigma^4\over4\ell^4}\cot^2{\omega_Lt\over2}\right)\sin^2{\omega_Lt\over2}.
\eea
One finds the usual focusing and antifocusing phenomena at
$t=nT$ and $t=(n+\hf)T$, respectively, where $T=2\pi/\omega_L$.
The crossover is at
\be
\sigma_{cr}=\ell\sqrt{2\left|\tan{\omega_Lt\over2}\right|}.
\ee
The wave function renormalization constant is found to be
\be
Z=4\pi\sigma^2.
\ee
One can see that the gauge invariant absolute magnitude of the wave
function preserves translation invariance and the norm of the blocked wave 
function follows the $\sigma$ dependence of the free particle. Nevertheless
the harmonic oscillator implicit in the system makes its appearance in 
$\sigma$ and the time dependence of the Gaussian peak of the absolute magnitude. 

{\it Hamiltonian:}
The gauge invariant generalization of blocking (\ref{hambl}), by using
\be
B(\sigma^2)=e^{-{\sigma^2_+\over2\hbar^2}(\s p-{e\over c}\s A)^2},\label{gib}
\ee
yields scale invariant hamiltonian as in the case of the free particle.

{\it Action:} The straightforward integration yields
\bea
{1\over\hbar}S_\sigma(\s u,\s v;t)&=&
{1\over2\tell^2}(\s u-\s v)^2+{1\over2\ell^2}(u_x+v_x)(u_y-v_y)\nonu
&&+i\log\left[4\pi i\ell^2\left(1-{\sigma^4\over4\ell^4}\right)
\left|\sin{\omega_Lt\over2}\right|\right].
\eea
The homogeneity of the magnetic field protects the action against picking up
scale dependence during the blocking as in the case of the free particle and the
only $\sigma$ dependence is in the additive constant. Notice that
this result is not as trivial as the scale invariance of the hamiltonian
because the gauge invariant smearing (\ref{gism}) does not corresponds to
(\ref{hambl}) and (\ref{gib}). The difference allows us to detect the
harmonic oscillator of the system by the singular scale dependence
of the additive constant in the blocked action.

The distinguished feature of systems with quadratic hamiltonians is 
that the imaginary part of the quantum action, $S(x,y;t)$ of (\ref{action}), 
is independent of the coordinates and the probability density
\be
\rho(x)=\left|e^{{i\over\hbar}S(x,y;\Delta t)}\right|^2
\ee
is $x$ and $y$ independent. This is a rather surprising result, it means for example
that a particle which is perfectly localized at the bottom of a harmonic potential
will be found any time later anywhere with homogeneous probability distribution.
Such an unphysical spread of the wave function is due to the perfect
localization (\ref{inic}). If one spreads the states by means of blocking (\ref{blwf}),
then the interference will introduce a realistic propagation pattern 
without invoking relativistic or multiparticle effects. The
essential modification of the dynamics by this spreading is indicated by the drastic
change of the propagation for weak smearing, $\sigma\to0$, c.f. (\ref{uvsing}).
But the nonharmonic terms of the hamiltonian strongly influence the probability distribution
and may alone remove these singularities. When the propagation between states with the same 
smearing is considered then the singularity at $\sigma\to0$ disappears.

\section{Representations of the Renormalization\break Group}
We study in this Section the renormalized trajectory 
\be
H(x,p)\longrightarrow H_\sigma(x,p)=e^{-{\sigma^2_+\over2\hbar^2}p^2}H(x,p)
e^{{\sigma^2_-\over2\hbar^2}p^2}
\ee
for the hamiltonians which are the sum of  the products of the canonical operators 
$x$ and $p$. We are interested in the quantum corrections to the evolution
equations. Thus we do not perform the trivial rescaling of the operators
by the observational length scale and we have the second term only
in the right hand side of (\ref{rgecf}). The simplest nontrivial case is
\be
x\longrightarrow x_\sigma=e^{-{\sigma^2_+\over2\hbar^2}p^2}x
e^{{\sigma^2_-\over2\hbar^2}p^2}=x+i\kappa p,
\ee
where
\be
\kappa={\sigma^2\over\hbar}.
\ee
This implies the transformation rule
\be
H(x,p)\longrightarrow H_\sigma(x,p)=H(x+i\kappa p,p).
\ee

For each n long string made up by the operators x and p, we associate an 
n-index tensor. The tensor indices take the values 1 and 2 and the only
nonvanishing matrix element is where the value of the j-th index is 1 (2) 
when the j-th operator in the product is x (p). The value of this matrix element
is the c-number coefficient of the product of $x$ and $p$. In this manner
the nonvanishing matrix elements of the tensor corresponding for
example to the operator $xxpx-2pxpx$ are $T^{1121}=1$ and $T^{2121}=-2$. 
The blocking is then represented by the linear transformation
\be
T^{a_1,\cdots,a_n}\longrightarrow g^{a_1,b_1}\cdots g^{a_n,b_n}T^{b_1,\cdots,b_n},
\label{actrg}
\ee
where
\be
g(\kappa)=\pmatrix{1&i\kappa\cr0&1}.\label{rgfr}
\ee
It is useful to write
\be
g(\kappa)=e^{i\kappa\sigma_+}=1+i\kappa\sigma_+,
\ee
where
\be
\sigma_+=\sigma_1+i\sigma_2=\pmatrix{0&1\cr0&0}.
\ee
The $g(\kappa)$ matrix gives the "fundamental representation" of the renormalization
group,
\be
g(\kappa_1)g(\kappa_2)=g(\kappa_1+\kappa_2)
\ee
what is a representation of a one dimensional subgroup of $sl(2,c)$.

The operator algebra generated by the products of $x$ and $p$ or the related
tensor algebra provides some linear representations of the renormalization group. 
But one should not forget that the coefficients of the operators, the matrix elements
of the tensors are nonlinear functions of the blocking parameter, $\sigma^2$.
It is obvious that the subspace $O_n$ which is span by the products of n operator 
of $x$ or $p$, i.e. the subspace of the n-index tensors, remains invariant
under the blocking.  Another invariant subspace
of the renormalization group $O_n(Y)$,
where $Y$ denotes an irreducible representation of the symmetric group
$S_n$, is span by such combination of products of n $x$ or $p$ operators which have the 
given symmetry with respect to the permutations of the operators in the products.
The antisymmetrisation with respect to the exchange of two $x$ or $p$ operators yields
0 or $i\hbar$. Thus the antisymmetrisation of an operator from $O_n$ yields
either another operator from $O_{n-2}$ or zero. The formal similarity of this
observation with the contraction of the indices by the help of the metric tensor
in the tensor representation of the group $sl(2,c)$ is evident. Thus it is enough
to study the subspaces $O_n(S)$ given in terms of the symmetrised 
Weyl-products. Since (\ref{rgfr}) is nonunitary, $O_n(S)$ is not necessarily
the direct sum of irreducible representations.

Let us consider the cases $n=2$ and 4 as examples where the most general
operator in $O_2(S)$ and $O_4(S)$ are
\bea
H_2&=&g_{2,0}h_{2,0}+ig_{1,1}h_{1,1}+g_{0,2}h_{0,2}\nonu
H_4&=&{1\over\sigma^2}\left(g_{4,0}h_{4,0}+ig_{3,1}h_{3,1}
+g_{2,2}h_{2,2}+ig_{1,3}h_{1,3}+g_{0,4}h_{0,4}\right),
\eea
where
\bea
h_{2,0}&=&p^2,\nonu
h_{1,1}&=&xp+px,\nonu
h_{0,2}&=&x^2,\nonu
h_{4,0}&=&p^4,\nonu
h_{3,1}&=&p^3x+p^2xp+pxp^2+xp^3,\nonu
h_{2,2}&=&p^2x^2+x^2p^2+xpxp+pxpx+xp^2x+px^2p,\nonu
h_{1,3}&=&px^3+xpx^2+x^2px+x^3p,\nonu
h_{0,4}&=&x^4.
\eea
By the help of the relations
\be
[e^{-{\sigma^2\over2\hbar^2}p^2},x^n]=
(\lambda h_{1,n-1}+\lambda^2h_{2,n-2}+\cdots\lambda^nh_{n,0})
 e^{-{\sigma^2\over2\hbar^2}p^2},
\ee
\be
[e^{-{\sigma^2\over2\hbar^2}p^2},h_{1,1}]=2\lambda h_{2,0}
 e^{-{\sigma^2\over2\hbar^2}p^2},
\ee
\be
[e^{-{\sigma^2\over2\hbar^2}p^2},h_{3,1}]=4\lambda h_{4,0}
 e^{-{\sigma^2\over2\hbar^2}p^2},
\ee
\be
[e^{-{\sigma^2\over2\hbar^2}p^2},h_{2,2}]=(3\lambda h_{3,1}
+6\lambda^2h_{4,0})e^{-{\sigma^2\over2\hbar^2}p^2},
\ee
\be
[e^{-{\sigma^2\over2\hbar^2}p^2},h_{1,3}]=(2\lambda h_{2,2}
+3\lambda^2h_{3,1}+4\lambda^3h_{4,0})e^{-{\sigma^2\over2\hbar^2}p^2},
\ee
where $\lambda=i\kappa$ we obtain the renormalization group transformation
\bea
g_{2,0}(\sigma^2)&=&g_{2,0}-2\kappa g_{1,1}-\kappa^2g_{0,2},\nonu
g_{1,1}(\sigma^2)&=&g_{1,1}+\kappa g_{0,2},\nonu
g_{0,2}(\sigma^2)&=&g_{0,2},\nonu
g_{4,0}(\sigma^2)&=&g_{4,0}-4\kappa g_{3,1}-6\kappa^2g_{2,2}
+4\kappa^3g_{1,3}+\kappa^4g_{0,4},\nonu
g_{3,1}(\sigma^2)&=&g_{3,1}+3\kappa g_{2,2}-3\kappa^2g_{1,3}-\kappa^3g_{0,4},\nonu
g_{2,2}(\sigma^2)&=&g_{2,2}-2\kappa g_{1,3}-\kappa^2g_{0,4},\nonu
g_{1,3}(\sigma^2)&=&g_{1,3}+\kappa g_{0,4},\nonu
g_{0,4}(\sigma^2)&=&g_{0,4}.
\eea
The nonvanishing of $g_{2,2}(\sigma^2)$ when $g_{0,4}(0)\not=0$ shows that the 
the anharmonic terms make the particle to appear as propagating on a 
curved manifold at the Landau pole, defined by $g_{2,0}(\sigma^2_L)=0$.

The renormalized trajectory for the harmonic oscillator is realized by $O_2(S)$
\be
H={p^2\over2m}+{m\omega^2\over2}x^2+ig(xp+px).
\ee
The only eigenvector is the hamiltonian of the free particle. The renormalization
group flow is
\be
\pmatrix{m^{-1}(\sigma)\cr\omega^2(\sigma)\cr g(\sigma)\cr}=\pmatrix{
m^{-1}\left(1-{\sigma^4\over\ell^4}\right)-4g{\sigma^2\over\hbar}\cr
\omega^2mm^{-1}(\sigma)\cr g+{m\omega^2\over2}{\sigma^2\over\hbar}}.
\ee
Bear in mind that $\omega(\sigma)$ is a parameter only, the true
frequency of the motion is left unchanged by blocking due to the
nonhermitean piece of the hamiltonian.

Our hamiltonian is ultraviolet finite so any point in the
coupling constant space $(m^{-1},\omega^2,g)$ is an ultraviolet fixed point.
The plane $\omega^2=0$, $\ell=\infty$ is invariant under the action of blocking. 
The flow with $\omega^2>0$ has more structure. In fact, the mass is not a 
monotonic function of $\sigma$ when $g<0$ and one can identify an 
ultraviolet and an infrared scaling region separated by a crossover
at $\sigma^2_{cr}=-2g\hbar/m\omega^2$ where the non-hermitean part of the 
hamiltonian changes sign. The usual harmonic oscillator is at the crossover 
and its evolution follows the $g>0$ infrared scaling regime.

$U_\ell\left(g(\kappa)\right)$, the $\ell$ angular momentum representation of 
(\ref{rgfr}) is an upper triangular matrix in the usual $|\ell,m\ra$ basis so
\be
{\rm det}\left[U_\ell\left(g(\kappa)\right)-\lambda\right]=(1-\lambda)^{2\ell+1}
\ee
and all eigenvalues of the representation of the renormalization group is 1.
One can verify easily that there is one eigenvector only in each space
$O_n(S)$ for even n, namely $h_{n,0}=p^n$. There is no eigenvector
for odd n as expected since the gaussian smearing cancels the
operators with odd space inversion parity. The Taylor expansion
of 
\be
U_\ell\left(g(\kappa)\right)=e^{i\kappa U_\ell(\sigma_+)}=
1+i\kappa U_\ell(\sigma_+)+\cdots
+{(                                                                                                                                                                                                                                       i\kappa)^{2\ell}\over2\ell!}U^{2\ell}_\ell(\sigma_+),
\ee
shows that an operator from $O_{2\ell+1}(S)$ is made of
$m+\ell$ $x$ operators where $-\ell\le m\le\ell$ contains the
$m+\ell$-th power of $\kappa$ in its transformation law under the renormalization
group transformation. Thus the operator with the largest weight in the infrared regime
will be $h_{n,0}=p^n$, $n=2\ell+1$.

\section{Summary}
A systematic method to incorporate in Quantum Mechanics,
the dependence of the dynamics on the
space resolution was presented in the spirit of the renormalization group.
The blocking step appears as the formal analogy of the Kadanoff-Wilson
blocking procedure of the second quantized quantum field variable repeated
on the level of the wave function. This was motivated by a simplified description
of the measurement process where the extended structure of the coherent part of
the measuring device was taken into account by the blocking of the wave function.
The scale dependence of the wave function and the dynamics extracted from it can be 
mapped out easily by borrowing ideas from the renormalization group.

The transformation of the operator algebra under blocking gives rise to a representation
of the renormalization group what happened to be a nonunitary representation
of a one dimensional subgroup of $sl(2,c)$. Several irreducible representations
were identified. It would be interesting to find all of them in order
to classify the possible "elementary dynamics". In the representations
provided by the homogeneous functions of the canonical variables, the 
momentum-dependent 
operators turned out to be the most relevant in the infrared limit. 
This is reasonable since on the one hand, the momentum-dependent terms
modify the propagation at arbitrary large distances and on the other hand,
the polynomial potentials generate bound states at finite energies
whose localization suppresses the infrared effects of the coordinate
dependent potential.

The smearing of the wave function is necessary in 
the case of the quadratic hamiltonains to arrive at acceptable transition 
amplitudes when a maximally localized Dirac-delta initial condition is
used. Three quadratic systems, the free particle, harmonic oscillator and a
two dimensional charged particle in homogeneous magnetic field are considered
in a detailed manner. The transition amplitudes from a maximally
localized state into a wave packet possess a crossover where the resolution
reaches the characteristic length scale of the problem. The blocked
dynamics of the free particle and the electric charge in a homogeneous magnetic
field is found to be scale invariant.

We finally note that the blocked propagator (\ref{genact}) provides a convenient
way to study the simultaneous dependence of the propagation both on the space resolution 
and on the observation time what might be useful in mesoscopic systems.

\section{Acknowledgement}
J.P. thanks Janos Hajdu for a useful discussion.

\end{document}